\documentclass[showpacs,preprintnumbers,twocolumn,superscriptaddress,aps,nofootinbib]{revtex4}
\usepackage{amssymb}
\usepackage[centertags]{amsmath}
\usepackage{txfonts}
\usepackage{epsfig}
\usepackage{bm}
\usepackage{color}
\usepackage{graphicx,graphics}
\usepackage{multirow}
\usepackage{float}
\usepackage{ulem}
\usepackage{appendix}
\usepackage[pdfstartview=FitH]{hyperref}
\hypersetup{colorlinks=true, citecolor=blue,linkcolor=blue,filecolor=black,urlcolor=blue}  
\allowdisplaybreaks[2]

\begin{document}

\title{Momentum dependence of light nuclei production in p-p, p-Pb and Pb-Pb collisions at the CERN Large Hadron Collider}

\author{Rui-Qin Wang}
\affiliation{School of Physics and Physical Engineering, Qufu Normal University, Shandong 273165, China}

\author{Feng-Lan Shao}
\email {shaofl@mail.sdu.edu.cn}
\affiliation{School of Physics and Physical Engineering, Qufu Normal University, Shandong 273165, China}

\author{Jun Song}
\email {songjun2011@jnxy.edu.cn}
\affiliation{Department of Physics, Jining University, Shandong 273155, China}

\begin{abstract}

We study the momentum dependence of the production of light nuclei in high energy collisions in the nucleon coalescence/recombination mechanism.
We derive formulas of the momentum distributions of deuterons ($d$) and helions ($^3$He).
We obtain the analytic expressions of the coalescence factor $B_A$'s ($B_2$ for $d$ and $B_3$ for $^3$He) as functions of the collision system size and the momentum.
We apply the deduced results to p-p, p-Pb and Pb-Pb collisions to naturally explain the interesting behaviors of $B_A$ observed in experiments at the CERN Large Hadron Collider.

\end{abstract}

\maketitle

\section{Introduction}

Light nuclei and anti-nuclei have provided a different window, compared to the electromagnetic and hadronic probes,
to study a series of fundamental problems in high energy physics~\cite{deuteron1960PRL,deuteron1961PRL,finalrecom1976PRL,QCDphase2010PLB,QCDphaseNuXu2020PRept,QCDphaseKLS2018PLB,QGP2019CERNYellow,dibaryon2019PRL,review2019NPA,Oliinychenko2021NPA}.
The study of the production of light (anti-)nuclei has attracted much attention in relativistic heavy ion collisions recently because they are considered to be effective tools to probe properties of the deconfined Quark Gluon Plasma and the Quantum Chromodynamics phase diagram ~\cite{QCDphaseKLS2018PLB,QGP2019CERNYellow,QCDphaseNuXu2020PRept,review2019NPA,Oliinychenko2021NPA}.
Two kinds of phenomenological models have proved to be particularly successful in describing the production of light nuclei.
One kind are specific models based on the coalescence/recombination mechanism~\cite{1coale1963PR,2coale1963PR,3coale1981PLB,4coale1991PRC,5coale1995PRL,6coale1996PRC,7coale1997PRC,8coale2003PRC,9WBZhao2018PRC,1BA1994PRL,2BA1998PLB,3BA1999PRC,4BA2018PRC,5BA2018MPLA,6BA2019PRC},
and the other are statistical models~\cite{1thermal1977PRL,2thermal1979PRL,3thermal2011PLB,4thermal2011PRC,5thermal2018nature,Tthermal2019PRC}.
Transport models such as those in Refs.~\cite{Danielewicz1991NPA,UrQMD2019PRC,Oliinychenko2019PRC,ART2009PRC,9BA2019PLG} have also been used to describe different production characteristics of light nuclei.

Experimental data on light nuclei production accumulated recently at the BNL Relativistic Heavy Ion Collider (RHIC) and the CERN Large Hadron Collider (LHC) exhibit a number of fascinating features~\cite{v2phid2007PRLPHENIX,v2d2016PRCSTAR,v1d2020PRCSTAR,DWZhang2021NPASTAR,B2B32001PRLSTAR,B2d2019PRCSTAR,v2PbPb2020PRCALICE,v2PbPb2020PLBALICE,ratioB22019NPAALICE,PbPb2016PRCALICE,PbPb2017EPJCALICE,pPb2020PLBALICE,pPb2020PRCALICE,pp2019PLBALICE,pp2020EPJCALICE}. 
The most striking ones might be the behaviors of the coalescence factor $B_A$ as functions of the size of the collision system and the transverse momentum per nucleon $p_T/A$~\cite{DWZhang2021NPASTAR,B2B32001PRLSTAR,B2d2019PRCSTAR,ratioB22019NPAALICE,PbPb2016PRCALICE,PbPb2017EPJCALICE,pPb2020PLBALICE,pPb2020PRCALICE,pp2019PLBALICE,pp2020EPJCALICE}.
The coalescence factor $B_A$ is defined as
\begin{eqnarray}
 E_A\frac{d^3N_A}{d p^3_A} = B_A \left( E_p\frac{d^3N_p}{d p^3_p} \right)^Z  \left( E_n\frac{d^3N_n}{d p^3_n} \right)^{A-Z},  \label{eq:BA}
\end{eqnarray}
where $E_Ad^3N_A/d p^3_A$ is the Lorentz-invariant momentum distribution of the light nuclei with mass number $A$ and charge $Z$, 
and $E_{p,n}d^3N_{p,n}/d p^3_{p,n}$ are those of protons and neutrons at momentum $p_{p,n}=p_A/A$.

From the viewpoint of coalescence/recombination mechanism, $B_A$ is a unique link between the formed light nuclei and the primordial nucleons. 
It carries the key kinetic and dynamical information of the process of nucleons coalescencing into light nuclei.
Moreover, $B_A$ can probe the freeze-out properties of the system such as the effective freeze-out volume~\cite{1BA1994PRL,YGMa2018PRept} and freeze-out particle correlations~\cite{6BA2019PRC} 
because nucleons recombining into light nuclei is expected to happen at a later stage of the system evolution~\cite{finalrecom1976PRL}.
Much effort has been put into the coalescence factor $B_A$ in different coalescence models~\cite{1BA1994PRL,2BA1998PLB,3BA1999PRC,4BA2018PRC,5BA2018MPLA,6BA2019PRC,7BA2019PRC,8BA2019APPB}.

In the year of 2020, the ALICE Collaboration just published more precise data of $B_A$ as the function of $p_T/A$ in both p-p and p-Pb collisions in separate intervals of multiplicity, which show nearly constant behaviors~\cite{pPb2020PLBALICE,pPb2020PRCALICE,pp2020EPJCALICE}.
In Pb-Pb collisions obvious increasing trend is observed for $B_A$ as the function of $p_T/A$~\cite{PbPb2016PRCALICE},
which is usually attributed to the position-momentum correlations or hard scatterings~\cite{2BA1998PLB,4BA2018PRC}.
A consistent and quantitive explanation for the different behaviors of the $p_T$ dependence of $B_A$ measured in different collision systems at the LHC is urgently necessary.

In this article, we apply the coalescence/recombination mechanism to hadronic systems created in p-p, p-nucleus (p-A) and nucleus-nucleus (A-A) collisions with extremely high collision energies 
to study the momentum dependence of light nuclei production in the low- and intermediate-$p_T$ regions.
We present simple formulas of momentum spectra and analytic expressions for momentum dependencies of $B_A$'s of different light nuclei.
We find that instantaneous coalescence/recombination occurring in the nucleon rest frame rather than in the laboratory frame can give natural explanations for the obvious growth of $B_A$ against $p_T$ for all centralities in Pb-Pb collisions and relatively weak $p_T$ dependence of $B_A$ in p-p and p-Pb collisions at the LHC~\cite{PbPb2016PRCALICE,PbPb2017EPJCALICE,pPb2020PLBALICE,pPb2020PRCALICE,pp2019PLBALICE,pp2020EPJCALICE}.

The rest of the paper is organized as follows. 
In Sec.~II, we present the derivation of the momentum dependence of the production of light nuclei in the framework of the nucleon coalescence/recombination. 
We present in particular the analytic expressions for the coalescence factors $B_2$ and $B_3$ measured by RHIC and LHC experiments~\cite{DWZhang2021NPASTAR,B2B32001PRLSTAR,B2d2019PRCSTAR,ratioB22019NPAALICE,PbPb2016PRCALICE,PbPb2017EPJCALICE,pPb2020PLBALICE,pPb2020PRCALICE,pp2019PLBALICE,pp2020EPJCALICE}
and discuss their qualitative properties. 
In Sec.~III, we apply the deduced results to p-p, p-Pb and Pb-Pb collisions at the LHC.
In Sec.~IV, we give a summary. 

%
\section{Formulas of light nuclei production in the coalescence/recombination mechanism}   \label{model}

In this section we start from the basic ideas of the coalescence/recombination and present general formulas of momentum dependence of light nuclei. 
Then we present the results obtained with several assumptions and/or approximations, such as the factorization of coordinate and momentum,
and those in modeling normalized nucleon coordinate distribution.
Finally, we give analytic results of coalescence factors $B_2$ and $B_3$ as functions of the system size and the momentum of light nuclei, and discuss their features. 

\subsection{The general formalism}  

We start with a hadronic system produced at the final stage of the evolution of high energy collision and suppose light nuclei are formed via the nucleon coalescence.
The three-dimensional momentum distribution of the produced deuterons $f_{d}(\bm{p})$ and that of helions $f_{^3\text{He}}(\bm{p})$ are given by
{\setlength\arraycolsep{0pt}
\begin{eqnarray}
 f_{d}(\bm{p})&=&  \int d\bm{x}_1d\bm{x}_2 d\bm{p}_1 d\bm{p}_2  f_{pn}(\bm{x}_1,\bm{x}_2;\bm{p}_1,\bm{p}_2)  \nonumber  \\
   &&~~~~ \times  \mathcal {R}_{d}(\bm{x}_1,\bm{x}_2;\bm{p}_1,\bm{p}_2,\bm{p}),      \label{eq:fdgeneral}   \\
 f_{^3\text{He}}(\bm{p})&=&  \int d\bm{x}_1d\bm{x}_2d\bm{x}_3 d\bm{p}_1 d\bm{p}_2 d\bm{p}_3  f_{ppn}(\bm{x}_1,\bm{x}_2,\bm{x}_3;\bm{p}_1,\bm{p}_2,\bm{p}_3) \nonumber  \\
   &&~~~~ \times \mathcal {R}_{^3\text{He}}(\bm{x}_1,\bm{x}_2,\bm{x}_3;\bm{p}_1,\bm{p}_2,\bm{p}_3,\bm{p}),      \label{eq:fHegeneral} 
\end{eqnarray} }%
where $f_{pn}(\bm{x}_1,\bm{x}_2;\bm{p}_1,\bm{p}_2)$ and $f_{ppn}(\bm{x}_1,\bm{x}_2,\bm{x}_3;\bm{p}_1,\bm{p}_2,\bm{p}_3)$ are two- and three- nucleon joint coordinate-momentum distributions, respectively; 
$\mathcal {R}_{d}(\bm{x}_1,\bm{x}_2;\bm{p}_1,\bm{p}_2,\bm{p})$ and $\mathcal {R}_{^3\text{He}}(\bm{x}_1,\bm{x}_2,\bm{x}_3;\bm{p}_1,\bm{p}_2,\bm{p}_3,\bm{p})$ are kernel functions.
Here and from now on we use bold symbols to denote three-dimensional coordinates and momenta.

The joint coordinate-momentum distributions $f_{pn}(\bm{x}_1,\bm{x}_2;\bm{p}_1,\bm{p}_2)$
and $f_{ppn}(\bm{x}_1,\bm{x}_2,\bm{x}_3;\bm{p}_1,\bm{p}_2,\bm{p}_3)$ are the nucleon number densities.
They satisfy 
{\setlength\arraycolsep{0pt}
\begin{eqnarray}
&& \int f_{pn}(\bm{x}_1,\bm{x}_2;\bm{p}_1,\bm{p}_2)  d\bm{x}_1d\bm{x}_2 d\bm{p}_1 d\bm{p}_2 = N_{pn},      \label{eq:fpnNpn}   \\
&& \int f_{ppn}(\bm{x}_1,\bm{x}_2,\bm{x}_3;\bm{p}_1,\bm{p}_2,\bm{p}_3) d\bm{x}_1d\bm{x}_2d\bm{x}_3 d\bm{p}_1 d\bm{p}_2 d\bm{p}_3\nonumber  \\ 
&&~~~~~~~~~~~~~~~~~~~~~~~~~~~~~~~~~~~~~~~~~~~~~~~~~~~~~~~~~~~~~~~~~~~~~ = N_{ppn},~~~~     \label{eq:fppnNppn} 
\end{eqnarray} }%
where $N_{pn}=N_{p}N_{n}$ and $N_{ppn}=N_{p}(N_{p}-1)N_{n}$ are the number of all possible $pn$-pairs and that of all possible $ppn$-clusters, respectively.
$N_{p}$ is the number of protons and $N_{n}$ is that of neutrons in the considered hadronic system.
We can rewrite the joint distributions in terms of the normalized forms denoted by the superscript $(n)$ as
{\setlength\arraycolsep{0pt}
\begin{eqnarray}
&& f_{pn}(\bm{x}_1,\bm{x}_2;\bm{p}_1,\bm{p}_2) = N_{pn} f^{(n)}_{pn}(\bm{x}_1,\bm{x}_2;\bm{p}_1,\bm{p}_2),      \label{eq:fpnNorm}   \\
&& f_{ppn}(\bm{x}_1,\bm{x}_2,\bm{x}_3;\bm{p}_1,\bm{p}_2,\bm{p}_3) = N_{ppn}   \nonumber \\  
&&~~~~~~~~~~~~~~~~~~~~~~~~~~~~~~~  \times  f^{(n)}_{ppn}(\bm{x}_1,\bm{x}_2,\bm{x}_3;\bm{p}_1,\bm{p}_2,\bm{p}_3). \label{eq:fppnNorm} 
\end{eqnarray} }%

Kernel functions $\mathcal {R}_{d}(\bm{x}_1,\bm{x}_2;\bm{p}_1,\bm{p}_2,\bm{p})$ and $\mathcal {R}_{^3\text{He}}(\bm{x}_1,\bm{x}_2,\bm{x}_3;\bm{p}_1,\bm{p}_2,\bm{p}_3,\bm{p})$ 
denote the probability density for $p,n$ with momenta $\bm{p}_1$ and $\bm{p}_2$ at $\bm{x}_1$ and $\bm{x}_2$ to recombine into a $d$ of momentum $\bm{p}$,
and that for $p,p,n$ with momenta $\bm{p}_1$, $\bm{p}_2$ and $\bm{p}_3$ at $\bm{x}_1$, $\bm{x}_2$ and $\bm{x}_3$ to recombine into a $^3$He of momentum $\bm{p}$, respectively.
Just as discussed in Ref.~\cite{RQWang2019CPC}, they carry the kinetic and dynamical information of the nucleons recombining into light nuclei,
and their precise expressions should be constrained by such as the momentum conservation, constraints due to intrinsic quantum numbers e.g. spin, and so on.
To take these constraints into account explicitly, we rewrite them in the following forms
{\setlength\arraycolsep{0pt}
\begin{eqnarray}
&&  \mathcal {R}_{d}(\bm{x}_1,\bm{x}_2;\bm{p}_1,\bm{p}_2,\bm{p}) = g_d \mathcal {R}_{d}^{(x,p)}(\bm{x}_1,\bm{x}_2;\bm{p}_1,\bm{p}_2) \delta(\displaystyle{\sum^2_{i=1}} \bm{p}_i-\bm{p}),  \nonumber  \\ \label{eq:Rdfac}  \\
&&  \mathcal {R}_{^3\text{He}}(\bm{x}_1,\bm{x}_2,\bm{x}_3;\bm{p}_1,\bm{p}_2,\bm{p}_3,\bm{p}) = g_{^3\text{He}}  \nonumber  \\
&&~~~~~~~~~~~~~~~~ \times \mathcal {R}_{^3\text{He}}^{(x,p)}(\bm{x}_1,\bm{x}_2,\bm{x}_3;\bm{p}_1,\bm{p}_2,\bm{p}_3)   \delta(\displaystyle{\sum^3_{i=1}} \bm{p}_i-\bm{p}) ,      \label{eq:RHefac}  
\end{eqnarray} }%
where the spin degeneracy factors $g_d=3/4$ and $g_{^3\text{He}}=1/4$.
The Dirac $\delta$ functions guarantee the momentum conservation in the coalescence.
The remaining $\mathcal {R}_{d}^{(x,p)}(\bm{x}_1,\bm{x}_2;\bm{p}_1,\bm{p}_2)$ now stands for the probability of a $pn$-pair with momenta $\bm{p}_1$ and $\bm{p}_2$ at $\bm{x}_1$ and $\bm{x}_2$ to recombine into a composite $d$-like particle with any momentum and any spin quantum number and similarly for $\mathcal {R}_{^3\text{He}}^{(x,p)}(\bm{x}_1,\bm{x}_2,\bm{x}_3;\bm{p}_1,\bm{p}_2,\bm{p}_3)$.
They depend on the positions and momenta of the nucleons and should be determined by the dynamics in the coalescence process.

In this way, we obtain the momentum distribution of $d$ and that of $^3$He as
{\setlength\arraycolsep{0pt}
\begin{eqnarray}
 f_{d}(\bm{p})&=& g_d N_{pn} \int d\bm{x}_1d\bm{x}_2 d\bm{p}_1 d\bm{p}_2  f^{(n)}_{pn}(\bm{x}_1,\bm{x}_2;\bm{p}_1,\bm{p}_2)  \nonumber  \\
   && \times  \mathcal {R}^{(x,p)}_{d}(\bm{x}_1,\bm{x}_2;\bm{p}_1,\bm{p}_2) \delta(\displaystyle{\sum^2_{i=1}} \bm{p}_i-\bm{p}),      \label{eq:fdgeneral1}   \\
 f_{^3\text{He}}(\bm{p})&=& g_{^3\text{He}} N_{ppn} \int d\bm{x}_1d\bm{x}_2d\bm{x}_3 d\bm{p}_1 d\bm{p}_2 d\bm{p}_3    \nonumber  \\
   && \times  f^{(n)}_{ppn}(\bm{x}_1,\bm{x}_2,\bm{x}_3;\bm{p}_1,\bm{p}_2,\bm{p}_3) \nonumber  \\
   && \times \mathcal {R}^{(x,p)}_{^3\text{He}}(\bm{x}_1,\bm{x}_2,\bm{x}_3;\bm{p}_1,\bm{p}_2,\bm{p}_3)  \delta(\displaystyle{\sum^3_{i=1}} \bm{p}_i-\bm{p}).      \label{eq:fHegeneral1} 
\end{eqnarray} }%
Eqs.~(\ref{eq:fdgeneral1}) and (\ref{eq:fHegeneral1}) are the general formulas for studying momentum distributions of light nuclei based on the basic ideas of the coalescence/recombination mechanism.
More specific results can be obtained when special assumptions or approximations are made about the normalized nucleon joint distributions and/or the coordinate-momentum dependent kernel functions.

\subsection{Factorization of coordinate and momentum dependencies}

Generally, the coordinate and momentum dependencies of kernel functions and nucleon joint distributions are coupled to each other,
especially in relativistic heavy ion collisions where the collective expansion exists. 
The coupling intensities and its specific forms are probably different in different collision systems/centralities.
Such coupling effects on light nuclei production have been studied in Ref.~\cite{2BA1998PLB}.
In this article, we focus on common features for production mechanisms of light nuclei in p-p, p-A and A-A collisions,
and in particular for a consistent understanding for the collision system size and momentum dependencies of the coalescence factor $B_A$.
We try our best to derive production formulas analytically and present momentum and collision system dependencies of light nuclei more intuitively,
so we do not include such coordinate-momentum coupling.
We consider only this case where the coordinate and momentum dependencies of kernel functions are decoupled from each other, i.e.,
{\setlength\arraycolsep{0pt}
\begin{eqnarray}
&&  \mathcal {R}^{(x,p)}_{d}(\bm{x}_1,\bm{x}_2;\bm{p}_1,\bm{p}_2) = \mathcal {R}_{d}^{(x)}(\bm{x}_1,\bm{x}_2) \mathcal {R}_{d}^{(p)}(\bm{p}_1,\bm{p}_2),      \label{eq:Rdxpfac}  \\
&&  \mathcal {R}^{(x,p)}_{^3\text{He}}(\bm{x}_1,\bm{x}_2,\bm{x}_3;\bm{p}_1,\bm{p}_2,\bm{p}_3) = \mathcal {R}_{^3\text{He}}^{(x)}(\bm{x}_1,\bm{x}_2,\bm{x}_3) \nonumber  \\
&&~~~~~~~~~~~~~~~~~~~~~~~~~~~~~~~~~~~~~~~~~~~~~~ \times    \mathcal {R}_{^3\text{He}}^{(p)}(\bm{p}_1,\bm{p}_2,\bm{p}_3),      \label{eq:RHexpfac}  
\end{eqnarray} }%
and the normalized joint distributions of the nucleons are coordinate and momentum factorized as follows
{\setlength\arraycolsep{0pt}
\begin{eqnarray}
&& f^{(n)}_{pn}(\bm{x}_1,\bm{x}_2;\bm{p}_1,\bm{p}_2) = f^{(n)}_{pn}(\bm{x}_1,\bm{x}_2)   f^{(n)}_{pn}(\bm{p}_1,\bm{p}_2),  \label{eq:fpnfac}  \\
&& f^{(n)}_{ppn}(\bm{x}_1,\bm{x}_2,\bm{x}_3;\bm{p}_1,\bm{p}_2,\bm{p}_3) = f^{(n)}_{ppn}(\bm{x}_1,\bm{x}_2,\bm{x}_3)  \nonumber  \\
&&~~~~~~~~~~~~~~~~~~~~~~~~~~~~~~~~~~~~~~~~~~~~ \times  f^{(n)}_{ppn}(\bm{p}_1,\bm{p}_2,\bm{p}_3).  \label{eq:fppnfac}    
\end{eqnarray} }%
Substituting Eqs.~(\ref{eq:Rdxpfac})-(\ref{eq:fppnfac}) into Eqs.~(\ref{eq:fdgeneral1}) and (\ref{eq:fHegeneral1}), we have
{\setlength\arraycolsep{0.2pt}
\begin{eqnarray}
&& f_{d}(\bm{p})= g_{d} N_{pn} \int d\bm{x}_1d\bm{x}_2 f^{(n)}_{pn}(\bm{x}_1,\bm{x}_2) \mathcal {R}^{(x)}_{d}(\bm{x}_1,\bm{x}_2)   \nonumber   \\
&& ~~~ \times
 \int d\bm{p}_1d\bm{p}_2 f^{(n)}_{pn}(\bm{p}_1,\bm{p}_2) \mathcal {R}^{(p)}_{d}(\bm{p}_1,\bm{p}_2) \delta(\displaystyle{\sum^2_{i=1}} \bm{p}_i-\bm{p}),~~~~~~     \label{eq:fd}  \\
&& f_{^3\text{He}}(\bm{p}) = g_{^3\text{He}} N_{ppn}  \nonumber   \\
&& ~~~ \times \int d\bm{x}_1d\bm{x}_2d\bm{x}_3 f^{(n)}_{ppn}(\bm{x}_1,\bm{x}_2,\bm{x}_3) \mathcal {R}^{(x)}_{^3\text{He}}(\bm{x}_1,\bm{x}_2,\bm{x}_3)  \nonumber   \\
&& ~~~ \times
 \int d\bm{p}_1d\bm{p}_2d\bm{p}_3 f^{(n)}_{ppn}(\bm{p}_1,\bm{p}_2,\bm{p}_3) \mathcal {R}^{(p)}_{^3\text{He}}(\bm{p}_1,\bm{p}_2,\bm{p}_3)    \nonumber \\
&&~~~~~~~~~~~~~~~~~~~~~~~~~~~~~~ \times  \delta(\displaystyle{\sum^3_{i=1}} \bm{p}_i-\bm{p}) .  \label{eq:fHe}  
\end{eqnarray} }%
Eqs.~(\ref{eq:fd}) and (\ref{eq:fHe}) show that we can calculate momentum distributions of different light nuclei by integrating coordinates and momenta of nucleons, respectively.

We use $\mathcal {A}_d$ and $\mathcal {A}_{^3\text{He}}$ to denote the coordinate integral parts in Eqs.~(\ref{eq:fd}) and (\ref{eq:fHe}) as
{\setlength\arraycolsep{0pt}
\begin{eqnarray}
&& \mathcal {A}_d =  \int d\bm{x}_1d\bm{x}_2 f^{(n)}_{pn}(\bm{x}_1,\bm{x}_2) \mathcal {R}^{(x)}_{d}(\bm{x}_1,\bm{x}_2),     \label{eq:Ad}  \\
&& \mathcal {A}_{^3\text{He}} = \int d\bm{x}_1d\bm{x}_2d\bm{x}_3 f^{(n)}_{ppn}(\bm{x}_1,\bm{x}_2,\bm{x}_3) \mathcal {R}^{(x)}_{^3\text{He}}(\bm{x}_1,\bm{x}_2,\bm{x}_3),~~~~~~  \label{eq:AHe}       
\end{eqnarray} }%
and use $\mathcal {M}_{d}(\bm{p})$ and $\mathcal {M}_{^3\text{He}}(\bm{p})$ to denote the momentum integral parts as
{\setlength\arraycolsep{0pt}
\begin{eqnarray}
&& \mathcal {M}_d (\bm{p}) =  \int d\bm{p}_1d\bm{p}_2 f^{(n)}_{pn}(\bm{p}_1,\bm{p}_2) \mathcal {R}^{(p)}_{d}(\bm{p}_1,\bm{p}_2) \delta(\displaystyle{\sum^2_{i=1}} \bm{p}_i-\bm{p}),  \nonumber  \\     \label{eq:Md}  \\
&& \mathcal {M}_{^3\text{He}} (\bm{p}) = \int d\bm{p}_1d\bm{p}_2d\bm{p}_3 f^{(n)}_{ppn}(\bm{p}_1,\bm{p}_2,\bm{p}_3) \mathcal {R}^{(p)}_{^3\text{He}}(\bm{p}_1,\bm{p}_2,\bm{p}_3) \nonumber  \\ 
&&~~~~~~~~~~~~~~~~~~~~~~~~~~~~~~~~~ \times  \delta(\displaystyle{\sum^3_{i=1}} \bm{p}_i-\bm{p}).  \label{eq:MHe}       
\end{eqnarray} }%
So we get
{\setlength\arraycolsep{0.2pt}
\begin{eqnarray}
&& f_{d}(\bm{p}) = g_{d} N_{pn} \mathcal {A}_d  \mathcal {M}_{d}(\bm{p}),     \label{eq:fd-AM}  \\
&& f_{^3\text{He}}(\bm{p}) = g_{^3\text{He}} N_{ppn} \mathcal {A}_{^3\text{He}}   \mathcal {M}_{^3\text{He}}(\bm{p}).     \label{eq:fHe-AM}  
\end{eqnarray} }%
$\mathcal {A}_d$ stands for the probability of a $pn$- pair satisfying the coordinate requirement to recombine into a deuteron-like molecular state,
and $\mathcal {M}_{d}(\bm{p})$ stands for the probability of a $pn$- pair satisfying the momentum requirement to recombine into a deuteron-like molecular state with momentum $\bm{p}$.
The similar case holds for $\mathcal {A}_{^3\text{He}}$ and $\mathcal {M}_{^3\text{He}}(\bm{p})$.
To evaluate these $\mathcal {A}_{d/^3\text{He}}$ and $\mathcal {M}_{d/^3\text{He}}(\bm{p})$, we need the coordinate and momentum factorized kernel functions and nucleon distributions.
In the following, we model their precise forms and give the corresponding analytic formulas for momentum distributions of light nuclei.

\subsection{Precise forms of kernel functions}

With the instantaneous coalescence approximation, kernel functions can be solved from the Wigner transformation once the wave functions of the light nuclei are given.
Just as in coalescence models in Refs.~\cite{Wigner2003NPA,Wigner2015PRC}, we adopt the wave function of a spherical harmonic oscillator 
and get the coordinate and momentum factorized Gaussian forms for our kernel functions.
Note that kernel functions should be directly determined by the relative coordinates and relative momenta of the primordial nucleons in the $pn$-pair (or $ppn$-cluster) rest frame rather than those in the laboratory frame~\cite{finalrecom1976PRL}. 
So we have
{\setlength\arraycolsep{0pt}
\begin{eqnarray}
&&  \mathcal {R}^{(x)}_{d}(\bm{x}_1,\bm{x}_2) = \mathcal {R}^{(x)}_{d}(\bm{x}'_1,\bm{x}'_2) = 8e^{-\frac{(\bm{x}'_1-\bm{x}'_2)^2}{\sigma_d^2}},      \label{eq:Rdx}  \\
&&  \mathcal {R}^{(p)}_{d}(\bm{p}_{1},\bm{p}_{2}) = \mathcal {R}^{(p)}_{d}(\bm{p}'_{1},\bm{p}'_{2})= e^{-\frac{\sigma_d^2(\bm{p}'_{1}-\bm{p}'_{2})^2}{4\hbar^2c^2}}, \label{eq:Rdp}  \\
&&  \mathcal {R}^{(x)}_{^3\text{He}}(\bm{x}_1,\bm{x}_2,\bm{x}_3) = \mathcal {R}^{(x)}_{^3\text{He}}(\bm{x}'_1,\bm{x}'_2,\bm{x}'_3)  \nonumber  \\
 &&~~~~~~~~~~~~~~~~~~~~~~~~~ =8^2e^{-\frac{(\bm{x}'_1-\bm{x}'_2)^2}{2\sigma_{^3\text{He}}^2}} e^{-\frac{(\bm{x}'_1+\bm{x}'_2-2\bm{x}'_3)^2}{6\sigma_{^3\text{He}}^2}},      \label{eq:RHex}   \\
&&  \mathcal {R}^{(p)}_{^3\text{He}}(\bm{p}_1,\bm{p}_2,\bm{p}_3) = \mathcal {R}^{(p)}_{^3\text{He}}(\bm{p}'_1,\bm{p}'_2,\bm{p}'_3)   \nonumber  \\
&&~~~~~~~~~~~~~~~~~~~~~~~~~=e^{-\frac{\sigma_{^3\text{He}}^2(\bm{p}'_{1}-\bm{p}'_{2})^2}{2\hbar^2c^2}} e^{-\frac{\sigma_{^3\text{He}}^2(\bm{p}'_{1}+\bm{p}'_{2}-2\bm{p}'_{3})^2}{6\hbar^2c^2}}.  \label{eq:RHep}
\end{eqnarray} }%
Here, as well as in the following of this article, the superscript `$'$' in the coordinate or momentum variable denotes the coordinate or momentum of the nucleon in the rest frame of the $pn$-pair or $ppn$-cluster.
The width parameter $\sigma_d=\sqrt{\frac{8}{3}} R_d$ and $\sigma_{^3\text{He}}=R_{^3\text{He}}$,
where $R_d$ and $R_{^3\text{He}}$ are the root-mean-square radius of the deuteron and that of the $^3$He, respectively.
The factor $\hbar c$ comes from the used GeV$\cdot$fm unit, and it is 0.197 GeV$\cdot$fm.

\subsection{Modeling the normalized coordinate distribution}

In this subsection we calculate $\mathcal {A}_d$ and $\mathcal {A}_{^3\text{He}}$.
Changing coordinate integral variables in Eq.~(\ref{eq:Ad}) to be $\bm{X}_C= \frac{\bm{x}_1+\bm{x}_2}{2}$ and $\bm{r}= \bm{x}_1-\bm{x}_2$, 
and those in Eq.~(\ref{eq:AHe}) to be $\bm{Y}_C= (\bm{x}_1+\bm{x}_2+\bm{x}_3)/\sqrt{3}$, $\bm{r}_1= (\bm{x}_1-\bm{x}_2)/\sqrt{2}$ and $\bm{r}_2= (\bm{x}_1+\bm{x}_2-2\bm{x}_3)/\sqrt{6}$,
and recalling Eqs.~(\ref{eq:Rdx}) and (\ref{eq:RHex}) we have
{\setlength\arraycolsep{0pt}
\begin{eqnarray}
&& \mathcal {A}_d =  8\int d\bm{X}_C d\bm{r} f^{(n)}_{pn}(\bm{X}_C,\bm{r}) e^{-\frac{\bm{r}'^2}{\sigma_d^2}},   \label{eq:Adr}   \\
&& \mathcal {A}_{^3\text{He}} =  8^2\int d\bm{Y}_Cd\bm{r}_1d\bm{r}_2 f^{(n)}_{ppn}(\bm{Y}_C,\bm{r}_1,\bm{r}_2) 
e^{-\frac{(\bm{r}'_1)^2+(\bm{r}'_2)^2}{\sigma_{^3\text{He}}^2}}. ~~~~  \label{eq:AHer}  
\end{eqnarray} }%
The normalization conditions become to be
{\setlength\arraycolsep{0pt}
\begin{eqnarray}
&& \int f^{(n)}_{pn}(\bm{X}_C,\bm{r})  d\bm{X}_Cd\bm{r}=1.  \\
&& \int f^{(n)}_{ppn}(\bm{Y}_C,\bm{r}_1,\bm{r}_2) d\bm{Y}_Cd\bm{r}_1d\bm{r}_2=1.
\end{eqnarray} }%
We further assume the coordinate joint distributions are coordinate variable factorized, i.e.,
$f^{(n)}_{pn}(\bm{X}_C,\bm{r}) = f^{(n)}_{pn}(\bm{X}_C) f^{(n)}_{pn}(\bm{r})$ and
$f^{(n)}_{ppn}(\bm{Y}_C,\bm{r}_1,\bm{r}_2) = f^{(n)}_{ppn}(\bm{Y}_C) f^{(n)}_{ppn}(\bm{r}_1) f^{(n)}_{ppn}(\bm{r}_2)$. 
Then we have
{\setlength\arraycolsep{0pt}
\begin{eqnarray}
&& \mathcal {A}_d =  8\int d\bm{r} f^{(n)}_{pn}(\bm{r}) e^{-\frac{\bm{r}'^2}{\sigma_d^2}} ,   \label{eq:Ad-initial}  \\
&& \mathcal {A}_{^3\text{He}} = 8^2 \int  d\bm{r}_1d\bm{r}_2 f^{(n)}_{ppn}(\bm{r}_1)  f^{(n)}_{ppn}(\bm{r}_2)
e^{-\frac{(\bm{r}'_1)^2+(\bm{r}'_2)^2}{\sigma_{^3\text{He}}^2}}.   \label{eq:AHe-initial} 
\end{eqnarray} }%

As in Ref.~\cite{fr2017acta}, we adopt $f^{(n)}_{pn}(\bm{r}) = \frac{1}{(\pi C R_f^2)^{1.5}} e^{-\frac{\bm{r}^2}{C R_f^2}}$ and
$f^{(n)}_{ppn}(\bm{r}_1) = \frac{1}{(\pi C_1 R_f^2)^{1.5}} e^{-\frac{\bm{r}_1^2}{C_1 R_f^2}}$, $f^{(n)}_{ppn}(\bm{r}_2) = \frac{1}{(\pi C_2 R_f^2)^{1.5}} e^{-\frac{\bm{r}_2^2}{C_2 R_f^2}}$,
where $R_f$ is the effective radius of the source system at the light nuclei freeze-out and $C$, $C_1$ and $C_2$ are distribution width parameters. 
Considering relations between $\bm{r}$, $\bm{r}_1$ and $\bm{r}_2$ with $\bm{x}_1$, $\bm{x}_2$ and $\bm{x}_3$, $C_1$ should be equal to $C/2$ 
and $C_2$ should be equal to $2C/3$.
So there is only one distribution width parameter $C$ to be determined.
In this article we set it to be 4, the same as that in Ref.~\cite{fr2017acta}. 

The general Lorentz transformation gives $\bm{x}' = \bm{x} +(\gamma-1)\frac{\bm{x}\cdot \bm{\beta}}{\beta^2}\bm{\beta} -\gamma t\bm{\beta}$, 
and the inverse Lorentz transformation gives $\bm{x} = \bm{x}' +(\gamma-1)\frac{\bm{x}'\cdot \bm{\beta}}{\beta^2}\bm{\beta} +\gamma t'\bm{\beta}$, 
where $\gamma=1/\sqrt{1-\beta^2}$ and $\bm{\beta}$ is the velocity of the center of mass of the $pn$-pair or $ppn$-cluster in the Laboratory frame.
Considering instantaneous coalescence/recombination in the rest frame of $pn$-pair or $ppn$-cluster, i.e., $\Delta t'=0$, we get
\begin{eqnarray}
\bm{r} = \bm{r}' +(\gamma-1)\frac{\bm{r}'\cdot \bm{\beta}}{\beta^2}\bm{\beta}.
\end{eqnarray}
Substituting the above equation into Eqs.~(\ref{eq:Ad-initial}) and (\ref{eq:AHe-initial}) and integrating from relative coordinate variables, we can obtain 
{\setlength\arraycolsep{0pt}
\begin{eqnarray}
 \mathcal {A}_{d} &=& \frac{8\sigma_d^3}{(C R_f^2+\sigma_d^2) \sqrt{C (R_f/\gamma)^2+\sigma_d^2}},   \label{eq:Ad-fin}  \\
 \mathcal {A}_{^3\text{He}} &=& \frac{8\sigma_{^3\text{He}}^3}{(\frac{C}{2} R_f^2+\sigma_{^3\text{He}}^2) \sqrt{\frac{C}{2} (R_f/\gamma)^2+\sigma_{^3\text{He}}^2}}  \nonumber  \\
&&  \times \frac{8\sigma_{^3\text{He}}^3}{(\frac{2C}{3} R_f^2+\sigma_{^3\text{He}}^2) \sqrt{\frac{2C}{3} (R_f/\gamma)^2+\sigma_{^3\text{He}}^2}} .   \label{eq:AHe-fin}  
\end{eqnarray} }%
The above two equations show that the coalescence probability in coordinate space becomes larger when applying coalescence criteria in the rest frame of the forming light nuclei compared to applying them in the laboratory frame where $\gamma=1$.
This can be equivalently understood as that the size of the fireball formed in the Laboratory is Lorentz contracted in the rest frame of the forming light nuclei, so the mean relative distance between nucleons becomes smaller and the coalescence becomes easier.

\subsection{Evaluating the nucleon momentum integral}

In this subsection, we evaluate $\mathcal {M}_d (\bm{p})$ and $\mathcal {M}_{^3\text{He}} (\bm{p})$.
Substituting Eqs.~(\ref{eq:Rdp}) and (\ref{eq:RHep}) into Eqs.~(\ref{eq:Md}) and (\ref{eq:MHe}), we have
{\setlength\arraycolsep{0.2pt}
\begin{eqnarray}
 \mathcal {M}_d (\bm{p}) &=&  \int d\bm{p}_1d\bm{p}_2 f^{(n)}_{pn}(\bm{p}_1,\bm{p}_2) e^{-\frac{\sigma_d^2(\bm{p}'_{1}-\bm{p}'_{2})^2}{4\hbar^2c^2}} \delta(\displaystyle{\sum^2_{i=1}} \bm{p}_i-\bm{p}),  \nonumber  \\     \label{eq:Md-initial}  \\
 \mathcal {M}_{^3\text{He}} (\bm{p}) &=&   \int d\bm{p}_1d\bm{p}_2d\bm{p}_3 f^{(n)}_{ppn}(\bm{p}_1,\bm{p}_2,\bm{p}_3) e^{-\frac{\sigma_{^3\text{He}}^2(\bm{p}'_{1}-\bm{p}'_{2})^2}{2\hbar^2c^2}} \nonumber  \\
&&~~~~~~ \times e^{-\frac{\sigma_{^3\text{He}}^2(\bm{p}'_{1}+\bm{p}'_{2}-2\bm{p}'_{3})^2}{6\hbar^2c^2}} \delta(\displaystyle{\sum^3_{i=1}} \bm{p}_i-\bm{p}) .     \label{eq:MHe-initial}  
\end{eqnarray} }%
Exact evaluations of $\mathcal {M}_d$ and $\mathcal {M}_{^3\text{He}}$ need the precise forms of nucleon joint momentum distributions which depend obviously on collision environments such as collision system, collision energy and collision centrality, etc. 
In this situation, the accurately analytic results for integrals in Eqs. (\ref{eq:Md-initial}) and (\ref{eq:MHe-initial}) are usually difficult to obtain. 
In order to get intuitionistic expressions for momentum dependence of light nuclei, in particular, those for $B_A$ factors, 
here we adopt the following mathematical approximation. 
Recalling that $\sigma_d=\sqrt{\frac{8}{3}} R_d$ and $\sigma_{^3\text{He}}=R_{^3\text{He}}$,
where the root-mean-square charge radius of the deuteron $R_d$=2.1421 fm and that of the $^3$He $R_{^3\text{He}}$=1.9661 fm \cite{radiiNPA2019},
we see that the gaussian width values $2\hbar c/\sigma_d$, $\sqrt{2}\hbar c/\sigma_{^3\text{He}}$ and $\sqrt{6}\hbar c/\sigma_{^3\text{He}}$ in Eqs.~(\ref{eq:Md-initial}) and (\ref{eq:MHe-initial}) are quite small.
So we can mathematically approximate the gaussian form of the kernel function $e^{-(\Delta \bm{p}')^2/\epsilon^2}$ as $(\sqrt{\pi} \epsilon)^3 \delta(\Delta \bm{p}')$, 
where $\epsilon$ is a small quantity.
Then we immediately obtain
{\setlength\arraycolsep{0.2pt}
\begin{eqnarray}
 \mathcal {M}_{d}(\bm{p}) 
&=&  (\frac{2\hbar c}{\sigma_d}\sqrt{\pi})^3
    \int d\bm{p}_1d\bm{p}_2 f^{(n)}_{pn}(\bm{p}_1,\bm{p}_2) \delta(\bm{p}'_{1}-\bm{p}'_{2})  \nonumber  \\
 && \times \delta(\displaystyle{\sum^2_{i=1}} \bm{p}_i-\bm{p}_d)    \nonumber   \\
 &=& (\frac{2\hbar c}{\sigma_d}\sqrt{\pi})^3 
    \int d\bm{p}_1d\bm{p}_2 f^{(n)}_{pn}(\bm{p}_1,\bm{p}_2) \gamma \delta(\bm{p}_{1}-\bm{p}_{2})  \nonumber  \\
 && \times \delta(\displaystyle{\sum^2_{i=1}} \bm{p}_i-\bm{p}_d)    \nonumber   \\
 &=& (\frac{\hbar c}{\sigma_d}\sqrt{\pi})^3 \gamma  f^{(n)}_{pn}(\frac{\bm{p}}{2},\frac{\bm{p}}{2}),  \label{eq:Md-mid}  
\end{eqnarray} }%
where $\gamma$ comes from $\Delta \bm{p}'=\frac{1}{\gamma}\Delta \bm{p}$.
Similarly we get
{\setlength\arraycolsep{0.2pt}
\begin{eqnarray}
 \mathcal {M}_{^3\text{He}} (\bm{p})
=  (\frac{\pi\hbar^2 c^2}{\sqrt{3}\sigma_{^3\text{He}}^2})^3 \gamma^2  f^{(n)}_{ppn}(\frac{\bm{p}}{3},\frac{\bm{p}}{3},\frac{\bm{p}}{3})  .     \label{eq:MHe-mid}  
\end{eqnarray} }%
Ignoring correlations between protons and neutrons, we finally have
{\setlength\arraycolsep{0.2pt}
\begin{eqnarray}
&& \mathcal {M}_{d}(\bm{p}) = (\frac{\hbar c}{\sigma_d}\sqrt{\pi})^3 \gamma  f^{(n)}_{p}(\frac{\bm{p}}{2}) f^{(n)}_{n}(\frac{\bm{p}}{2}) ,  \label{eq:Md-fin}  \\
&& \mathcal {M}_{^3\text{He}} (\bm{p}) =  (\frac{\pi\hbar^2 c^2}{\sqrt{3}\sigma_{^3\text{He}}^2})^3 \gamma^2  f^{(n)}_{p}(\frac{\bm{p}}{3}) 
 f^{(n)}_{p}(\frac{\bm{p}}{3})   f^{(n)}_{n}(\frac{\bm{p}}{3}) .     \label{eq:MHe-fin} 
\end{eqnarray} }%
To check the robustness of the above $\delta$ function approximation, we also take a classical Boltzmann distribution $f^{(n)}_{p,n}= \frac{1}{(2\pi mT)^{1.5}}e^{-\bm{p}^2/(2mT)}$ for nucleons as an example 
to carry out practical integrals in Eqs. (\ref{eq:Md-initial}) and (\ref{eq:MHe-initial}).
We can obtain the exact results of momentum integrals. 
Considering the relationship $mT\sigma_{d/^3\text{He}}^2\gg \gamma^2\hbar^2c^2$, 
we can express integrated results as Taylor series and the leading terms are just Eqs. (\ref{eq:Md-fin}) and (\ref{eq:MHe-fin}).

\subsection{Momentum distributions of light nuclei}

Substituting Eqs.~(\ref{eq:Ad-fin}-\ref{eq:AHe-fin}) and (\ref{eq:Md-fin}-\ref{eq:MHe-fin}) into Eqs.~(\ref{eq:fd-AM}) and (\ref{eq:fHe-AM}),
we finally have the momentum distributions of light nuclei as
{\setlength\arraycolsep{0.2pt}
\begin{eqnarray}
&& f_{d}(\bm{p}) = \frac{ 8(\sqrt{\pi}\hbar c)^3 g_{d} \gamma}{(C R_f^2+\sigma_d^2) \sqrt{C (R_f/\gamma)^2+\sigma_d^2}} 
                       f_{p}(\frac{\bm{p}}{2}) f_{n}(\frac{\bm{p}}{2}) ,     \label{eq:fd-final}  \\
&& f_{^3\text{He}}(\bm{p}) = \frac{8^2 (\pi\hbar^2 c^2)^3 g_{^3\text{He}} \gamma^2 }{3\sqrt{3}(\frac{C}{2} R_f^2+\sigma_{^3\text{He}}^2) \sqrt{\frac{C}{2} (R_f/\gamma)^2+\sigma_{^3\text{He}}^2} } \nonumber  \\
&&    \times  \frac{1}{(\frac{2C}{3} R_f^2+\sigma_{^3\text{He}}^2) \sqrt{\frac{2C}{3} (R_f/\gamma)^2+\sigma_{^3\text{He}}^2}} 
  f_{p}(\frac{\bm{p}}{3}) f_{p}(\frac{\bm{p}}{3}) f_{n}(\frac{\bm{p}}{3}). ~~~~~~    \label{eq:fHe-final}  
\end{eqnarray} }%
Eqs.~(\ref{eq:fd-final}) and (\ref{eq:fHe-final}) show relationships of light nuclei with primordial nucleons in momentum space in the laboratory frame.
They can be used to calculate the yields and transverse momentum spectra of light nuclei measured extensively as long as the the nucleon momentum distributions are given.
They can also be conveniently used to probe production correlations of light nuclei and nucleons, such as the coalescence factor which will be discussed in the next subsection.

\subsection{Coalescence factor $B_A$}

In this subsection, we explore analytic results of coalescence factors $B_A$'s.
Noting that in Eq.~(\ref{eq:fd-final}) $f_{d}(\bm{p}) = {d^3N_d}/{d \bm{p}_d}$ and $f_{p,n}(\bm{p}) = {d^3N_{p,n}}/{d \bm{p}_{p,n}}$, we have
{\setlength\arraycolsep{0.2pt}
\begin{eqnarray}
B_2 &\equiv& \left( E_d\frac{d^3N_d}{d \bm{p}_d} \right) / \left[ (E_p\frac{d^3N_p}{d \bm{p}_p}) (E_n\frac{d^3N_n}{d \bm{p}_n}) \right]   \nonumber     \\
    &=& \frac{ 32(\sqrt{\pi}\hbar c)^3 g_{d} }{m_d (C R_f^2+\sigma_d^2) \sqrt{C (R_f/\gamma)^2+\sigma_d^2}},  \label{eq:B2}
\end{eqnarray} }%
where $E_p=E_n=\frac{1}{2}E_d=\frac{1}{2}\gamma m_d$ is used for the second equality.
The mass of the deuteron is $m_d=1.875$ GeV and $\gamma=\sqrt{1+\bm{p}_d^2/m_d^2}$.
Our result Eq.~(\ref{eq:B2}) is consistent with previous works \cite{3BA1999PRC,6BA2019PRC} as one notices that we define $R_f$ in the Laboratory frame while Refs.~\cite{6BA2019PRC} and \cite{3BA1999PRC} defined these fireball radius parameters in the nucleon pair rest frame and Yano-Koonin-Podgoretski\u\i\ (YKP) frame, respectively.

Similarly for $^3$He, we have
{\setlength\arraycolsep{0.2pt}
\begin{eqnarray}
B_3
    &=& \frac{192 \sqrt{3} (\pi\hbar^2 c^2)^3 g_{^3\text{He}}  }{ m^2_{^3\text{He}} (\frac{C}{2} R_f^2+\sigma_{^3\text{He}}^2) \sqrt{\frac{C}{2} (R_f/\gamma)^2+\sigma_{^3\text{He}}^2} } \nonumber  \\
&&   ~ \times  \frac{1}{(\frac{2C}{3} R_f^2+\sigma_{^3\text{He}}^2) \sqrt{\frac{2C}{3} (R_f/\gamma)^2+\sigma_{^3\text{He}}^2}}.  \label{eq:B3}
\end{eqnarray} }%
The mass of $^3$He is $m_{^3\text{He}}=2.815$ GeV and $\gamma=\sqrt{1+\bm{p}_{^3\text{He}}^2/m_{^3\text{He}}^2}$.
Eqs.~(\ref{eq:B2}) and (\ref{eq:B3}) are the final results for $B_A$ factors we derived, 
which clearly show that $B_2$ and $B_3$ depend sensitively on the system size denoted by $R_f$, the size of light nuclei via $\sigma_{d/^3\text{He}}$ and the momentum of light nuclei via $\gamma$.

From Eqs.~(\ref{eq:B2}) and (\ref{eq:B3}), we can get properties of $B_A$ factors.
The first is the larger $R_f$, the smaller $B_2$ and $B_3$. 
This means $B_A$ always becomes smaller from small p-p collisions to semi-central Pb-Pb collisions and then to central Pb-Pb collisions.
The second is that finite sizes of light nuclei suppress their production, and the suppression is stronger in small p-p and p-Pb collisions than in Pb-Pb collisions.
The last but the most important, Eqs.~(\ref{eq:B2}) and (\ref{eq:B3}) give explicitly the momentum dependence of $B_2$ and $B_3$ via $\gamma$,
In the case of $R_f\gg\sigma_d$ or $\sigma_{^3\text{He}}$, such as in central Pb-Pb collisions at the LHC, both $B_2$ and $B_3$ increase with the increasing momentum.
In the case of $R_f\ll\sigma_d$ or $\sigma_{^3\text{He}}$, such as in small p-p collisions, $B_2$ and $B_3$ nearly keep invariant with the momentum.
All these properties of $B_A$ are characteristics of light nuclei production in the coalescence/recombination production mechanism. 
They can be directly tested by the experimental data.


\section{Applications in $\text{p-p}$, $\text{p-Pb}$ and $\text{Pb-Pb}$ collisions at the LHC} \label{application}

In this section, we apply the deduced results in Sec.~\ref{model} to the midrapidity region of p-p, p-Pb and Pb-Pb collisions at the LHC
to study behaviors of $B_A$ as functions of the collision system size and the transverse momentum of light nuclei. 
First we present results of $B_2$ and $B_3$ as the function of the effective radius of the hadronic system $R_f$.
Then we give results of $B_2$ and $B_3$ as the function of the charged particle pseudorapidity density ${dN_{ch}}/{d\eta}$. 
Finally we show results of $B_2$ and $B_3$ as the function of the transverse momentum per nucleon $p_T/A$.

\subsection{$B_A$ as the function of $R_f$} 

\begin{figure}[htbp]
\centering
 \includegraphics[width=1.\linewidth]{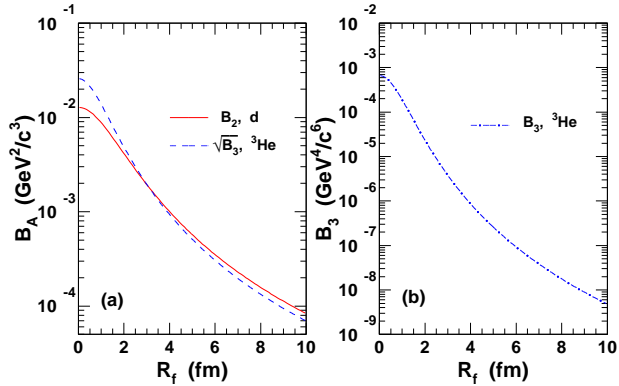}\\
 \caption{(Color online) (a) $B_2$ of $d$ and $\sqrt{B_3}$ of $^3$He, (b) $B_3$ of $^3$He as the function of $R_f$ at $p_T/A=0.75$ GeV/c. 
 }
 \label{fig:BA-Rf}
\end{figure}

With Eqs.~(\ref{eq:B2}) and (\ref{eq:B3}), we calculate $B_2$ of $d$ and $B_3$ of $^3$He at a fixed value of $p_T/A=0.75$ GeV/c as the function of the collision system size denoted by $R_f$.
The results are presented with the solid line and the dash-dotted line in Fig.~\ref{fig:BA-Rf} (a) and (b), respectively,
which show that both $B_2$ and $B_3$ decrease with the increase of $R_f$, and $B_3$ decreases more rapidly than $B_2$.

In large $R_f$ region where $R_f\gg\sigma_{d/^3\text{He}}$, Eqs.~(\ref{eq:B2}) and (\ref{eq:B3}) give
{\setlength\arraycolsep{0.2pt}
\begin{eqnarray}
B_2\varpropto R_f^{-3} \varpropto V_f^{-1},  \label{eq:B2Vf} \\
B_3\varpropto R_f^{-6} \varpropto V_f^{-2}.  \label{eq:B3Vf}
\end{eqnarray} }%
$V_f$ is the effective volume of the collision system at the freeze-out of light nuclei.
These results are consistent with those in Refs.~\cite{7BA2019PRC,8BA2019APPB}.
From Eqs.~(\ref{eq:B2Vf}) and (\ref{eq:B3Vf}) one can see $\sqrt{B_3}$ should have similar behaviors with that of $B_2$ as the function of $R_f$.
We plot the result of $\sqrt{B_3}$ with the dashed line in Fig.~\ref{fig:BA-Rf} (a),
and find it is indeed almost parallel to $B_2$ in large $R_f$ region such as $R_f>4$ fm.
But in small $R_f$ area, $\sqrt{B_3}$ is different from $B_2$,
because in this area they are affected by not only $R_f$ but also $\sigma_{d/^3\text{He}}$ related with different sizes of $d$ and $^3$He themselves.

\subsection{$B_A$ as the function of ${dN_{ch}}/{d\eta}$} 

\begin{figure}[htbp]
\centering
 \includegraphics[width=1.0\linewidth]{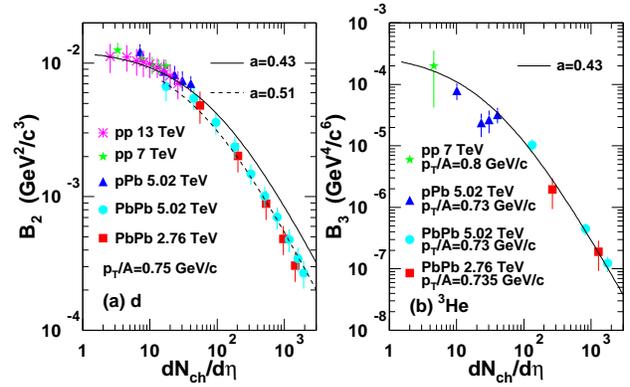}\\
 \caption{(Color online) (a) $B_2$ of $d$ and (b) $B_3$ of $^3$He as the function of $dN_{ch}/d\eta$ in p-p, p-Pb and  Pb-Pb collisions at the LHC. 
Filled symbols are the data~\cite{PbPb2016PRCALICE,pPb2020PLBALICE,pPb2020PRCALICE,pp2019PLBALICE,pp2020EPJCALICE,PbPb5TeVthesis}, and different lines are our results.
 }
 \label{fig:BA-dNch}
\end{figure}

Pseudorapidity density of the charged particles ${dN_{ch}}/{d\eta}$ is the most direct observable to represent the size of the collision system.
We in this subsection study $B_2$ of $d$ and $B_3$ of $^3$He as the function of ${dN_{ch}}/{d\eta}$.
To compare with the experimental data, we choose $p_T/A=0.75$ GeV/c for $d$ and $p_T/A=0.735$ GeV/c for $^3$He.
The relationship between ${dN_{ch}}/{d\eta}$ and $R_f$ can be parameterized based on the Hanbury-Brown-Twiss interferometry as $R_f=a*(dN_{ch}/d\eta)^{1/3}$~\cite{HBTreview2005}.
Here the proportionality coefficient $a$ is a free parameter, and it is located in the range 0.4-1.0 extracted from the HBT correlations in Ref.~\cite{HBT2016PRC}.
We here first adopt $a=0.43$ to compute $B_2$ and $B_3$.
Solid lines in Fig.~\ref{fig:BA-dNch} are our results and filled symbols are the experimental data~\cite{PbPb2016PRCALICE,pPb2020PLBALICE,pPb2020PRCALICE,pp2019PLBALICE,pp2020EPJCALICE,PbPb5TeVthesis}. 
Overall, our results agree with the data of $B_2$ and $B_3$ in different collision systems at LHC energies except overestimations of $B_2$ in Pb-Pb collisions.

We retune $a$ to be 0.51 to recompute $B_2$ in Pb-Pb collisions and the result presented by the dashed line in Fig.~\ref{fig:BA-dNch} (a) can reproduce
the data very well.
In our model, different value of $a$ means different $R_f$ at a given $dN_{ch}/d\eta$ and further means different freeze-out time.
0.51 for $d$ and 0.43 of $a$ for $^3$He mean that the effective radius of the system at $d$ freeze-out is about 20\% larger than that at $^3$He freeze-out.
This also implies later freeze out for $d$ compared to $^3$He in Pb-Pb collisions. 
We note that this is consistent with the work of the Blast-wave model where it gives that 
the kinetic freeze-out temperature (transverse expansion velocity) of $d$ is lower (larger) than that of $^3$He~\cite{PbPb2016PRCALICE}.
Richer measurements for $^3$He are expected to make precise conclusions.

\subsection{$B_A$ as the function of the transverse momentum}

\begin{figure*}[htbp]
\centering
 \includegraphics[width=0.8\linewidth]{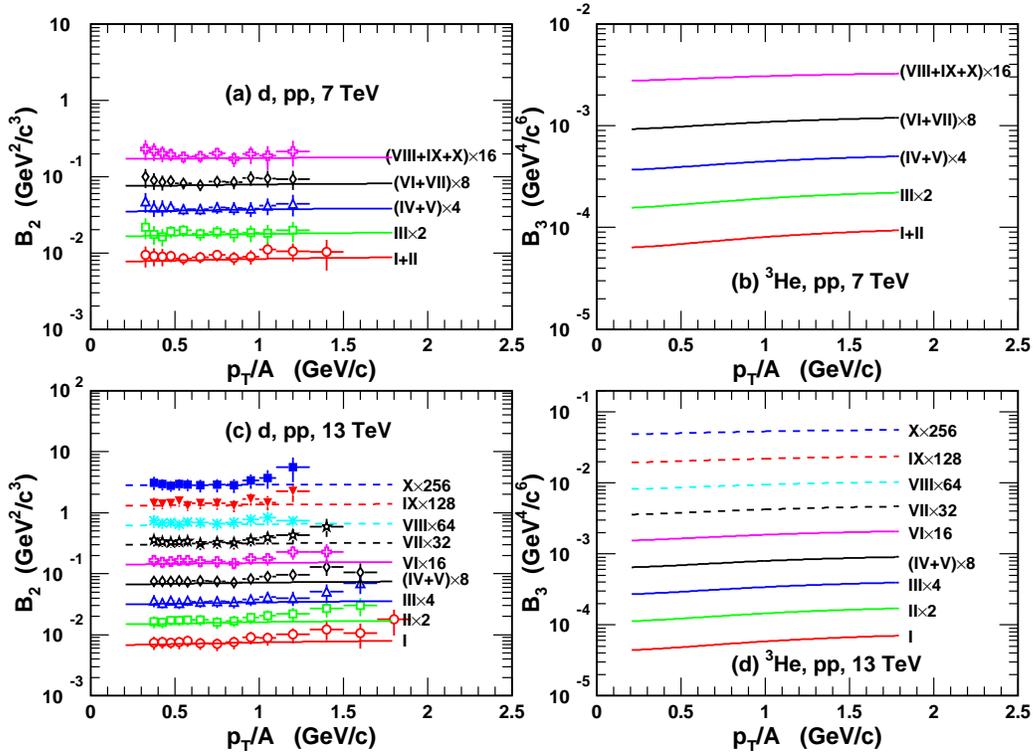}\\
 \caption{(Color online) (a) $B_2$ of deuterons and (b) $B_3$ of $^3$He as functions of $p_T/A$ in p-p collisions at $\sqrt{s}=7$ TeV. 
 (c) $B_2$ of deuterons and (d) $B_3$ of $^3$He as functions of $p_T/A$ in p-p collisions at $\sqrt{s}=13$ TeV.
Symbols are the data from Refs.~\cite{pp2019PLBALICE,pp2020EPJCALICE}, and different lines are our results.
 $B_2$ and $B_3$ have been scaled with the indicated factors for better visibility.
 }
 \label{fig:BA-ptApp7TeV13TeV}
\end{figure*}

In this subsection we study the transverse momentum dependence of $B_A$ in the midrapidity regions in p-p, p-Pb and Pb-Pb collisions at the LHC.
Recalling Eqs.~(\ref{eq:B2}) and (\ref{eq:B3}) and considering $\gamma=\sqrt{1+p_T^2/m^2_{d/^3\text{He}}}$ at the midrapidity, we have
{\setlength\arraycolsep{0.2pt}
\begin{eqnarray}
&& B_2(p_T) = \frac{ 32(\sqrt{\pi}\hbar c)^3 g_{d} \sqrt{1+\frac{p_T^2}{m^2_d}} }{m_d (C R_f^2+\sigma_d^2) \sqrt{C R_f^2+(1+\frac{p_T^2}{m^2_d})\sigma_d^2}},  \label{eq:B2pT}  \\
&&B_3(p_T)= \frac{192 \sqrt{3} (\pi\hbar^2 c^2)^3 g_{^3\text{He}} (1+\frac{p_T^2}{m^2_{^3\text{He}}}) }{ m^2_{^3\text{He}} (\frac{C}{2} R_f^2+\sigma_{^3\text{He}}^2) \sqrt{\frac{C}{2} R_f^2+(1+\frac{p_T^2}{m^2_{^3\text{He}}})\sigma_{^3\text{He}}^2} } \nonumber  \\
&& ~~~~~~ \times  \frac{1}{(\frac{2C}{3} R_f^2+\sigma_{^3\text{He}}^2) \sqrt{\frac{2C}{3} R_f^2+(1+\frac{p_T^2}{m^2_{^3\text{He}}})\sigma_{^3\text{He}}^2}}.  \label{eq:B3pT}
\end{eqnarray} }%
The above two equations show that $B_2$ and $B_3$ have close relations with the $p_T$ of light nuclei.

In particular, in the case that $R_f$ is much larger than sizes of light nuclei, such as in central Pb-Pb collisions at the LHC, 
we have 
{\setlength\arraycolsep{0.2pt}
\begin{eqnarray}
&& B_2(p_T) = \frac{ 32(\sqrt{\pi}\hbar c)^3 g_{d} }{ m_d C^{3/2} R_f^3 }   \sqrt{1+\frac{p_T^2}{m^2_d}},  \label{eq:B2pTLRf}  \\
&&B_3(p_T)= \frac{1728 (\pi\hbar^2 c^2)^3 g_{^3\text{He}} }{ m^2_{^3\text{He}} C^{3} R_f^6  }  (1+\frac{p_T^2}{m^2_{^3\text{He}}}).  \label{eq:B3pTLRf}
\end{eqnarray} }%
Eqs.~(\ref{eq:B2pTLRf}) and (\ref{eq:B3pTLRf}) show both $B_2$ and $B_3$ should increase with $p_T$, and such increase trend of $B_3$ is stronger than that of $B_2$.
Otherwise, in the limit case that $R_f$ is much smaller than sizes of light nuclei, we have
{\setlength\arraycolsep{0.2pt}
\begin{eqnarray}
&& B_2(p_T) = \frac{ 32(\sqrt{\pi}\hbar c)^3 g_{d} }{m_d \sigma_d^3 },  \label{eq:B2pTsRf}  \\
&&B_3(p_T)= \frac{192 \sqrt{3} (\pi\hbar^2 c^2)^3 g_{^3\text{He}}  }{ m^2_{^3\text{He}} \sigma_{^3\text{He}}^6  } .  \label{eq:B3pTsRf}
\end{eqnarray} }%
In this limit case, $B_2$ and $B_3$ are independent of $p_T$.
From the above discussions, one can see that $p_T$ dependence of $B_A$ is different in different collisions.
Such interesting behaviors of $B_A$'s as the function of $p_T$ in different collision systems are natural characteristics of our model.
They can be used to test the validity of our model and the production mechanisms of light nuclei.
Next we will test them in p-p, p-Pb and Pb-Pb collisions in LHC experiments.

We first study $B_2$ and $B_3$ as functions of $p_T/A$ in p-p collisions at $\sqrt{s}=7$ TeV and 13 TeV.
Results are in Fig.~\ref{fig:BA-ptApp7TeV13TeV}.
We use the data of $dN_{ch}/d\eta$ in Refs.~\cite{dNch7pp,dNch13pp} to determine $R_f$ by $R_f=a*(dN_{ch}/d\eta)^{1/3}$ with adopting $a=0.43$.
Symbols are the data from Refs.~\cite{pp2019PLBALICE,pp2020EPJCALICE}, and different lines are our results.
$B_2$ and $B_3$ have been scaled with the indicated factors for better visibility.
From Fig.~\ref{fig:BA-ptApp7TeV13TeV}, one can see that $B_2$ exhibits nearly constant trends within error bars and $B_3$ shows very weak $p_T$ dependence.
This is the natural result of Eqs.~(\ref{eq:B2pT}) and (\ref{eq:B3pT}).
In small p-p collisions, $R_f$ is even smaller than the sizes of the light nuclei themselves denoted by $\sigma_d$ or $\sigma_{^3\text{He}}$.
In this case, the item containing $1+\frac{p_T^2}{m^2_{d/^3\text{He}}}$ in the denominator becomes prominent and it can nearly offset the $p_T$ dependence in the numerator.
In the limit case of $R_f=0$ fm, $p_T$ dependence in the denominator can completely offset $p_T$ dependence in the numerator, and $B_A$ exhibits constant behavior as the function of $p_T/A$.
So the smaller of the collision system, the weaker $p_T$ dependence of $B_A$ due to the nonnegligible size of the light nuclei.

\begin{figure}[htbp]
\centering
 \includegraphics[width=1.\linewidth]{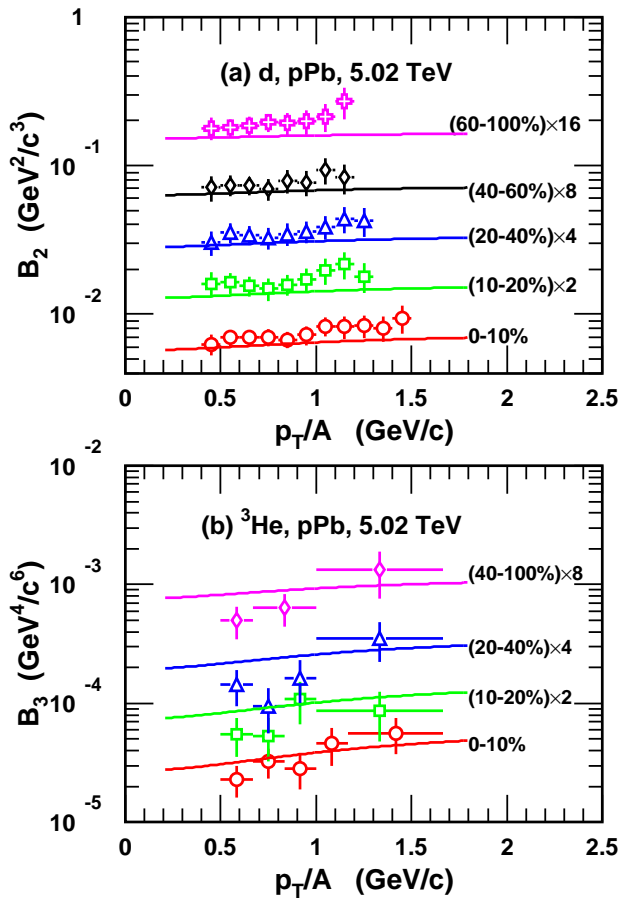}\\
 \caption{(Color online) (a) $B_2$ of deuterons and (b) $B_3$ of $^3$He as functions of $p_T/A$ in p-Pb collisions at $\sqrt{s_{NN}}=5.02$ TeV.
 Open symbols are the data~\cite{pPb2020PLBALICE,pPb2020PRCALICE}, and different lines are our results.
 $B_2$ and $B_3$ have been scaled with the indicated factors for better visibility.
 }
 \label{fig:BA-ptApPb}
\end{figure}

We then study $B_2$ and $B_3$ as functions of $p_T/A$ in p-Pb collisions at $\sqrt{s_{NN}}=5.02$ TeV.
Results are in Fig.~\ref{fig:BA-ptApPb}.
We use the data of $dN_{ch}/d\eta$ in Ref.~\cite{dNchpPb2014PLB} to determine $R_f$.
Here $a$ is 0.43, the same as that in p-p collisions.
Open symbols are the data~\cite{pPb2020PLBALICE,pPb2020PRCALICE}, and different lines are our results.
$B_2$ and $B_3$ have been scaled with the indicated factors for better visibility.
From Fig.~\ref{fig:BA-ptApPb}, one can see that our results can reproduce behaviors of $B_2$ and $B_3$ measured experimentally.
In p-Pb collisions, $R_f$ becomes to be comparable to the sizes of the light nuclei themselves.
So $B_2$ and $B_3$ begin to increase as the function of $p_T$.

\begin{figure*}[htbp]
\centering
 \includegraphics[width=0.9\linewidth]{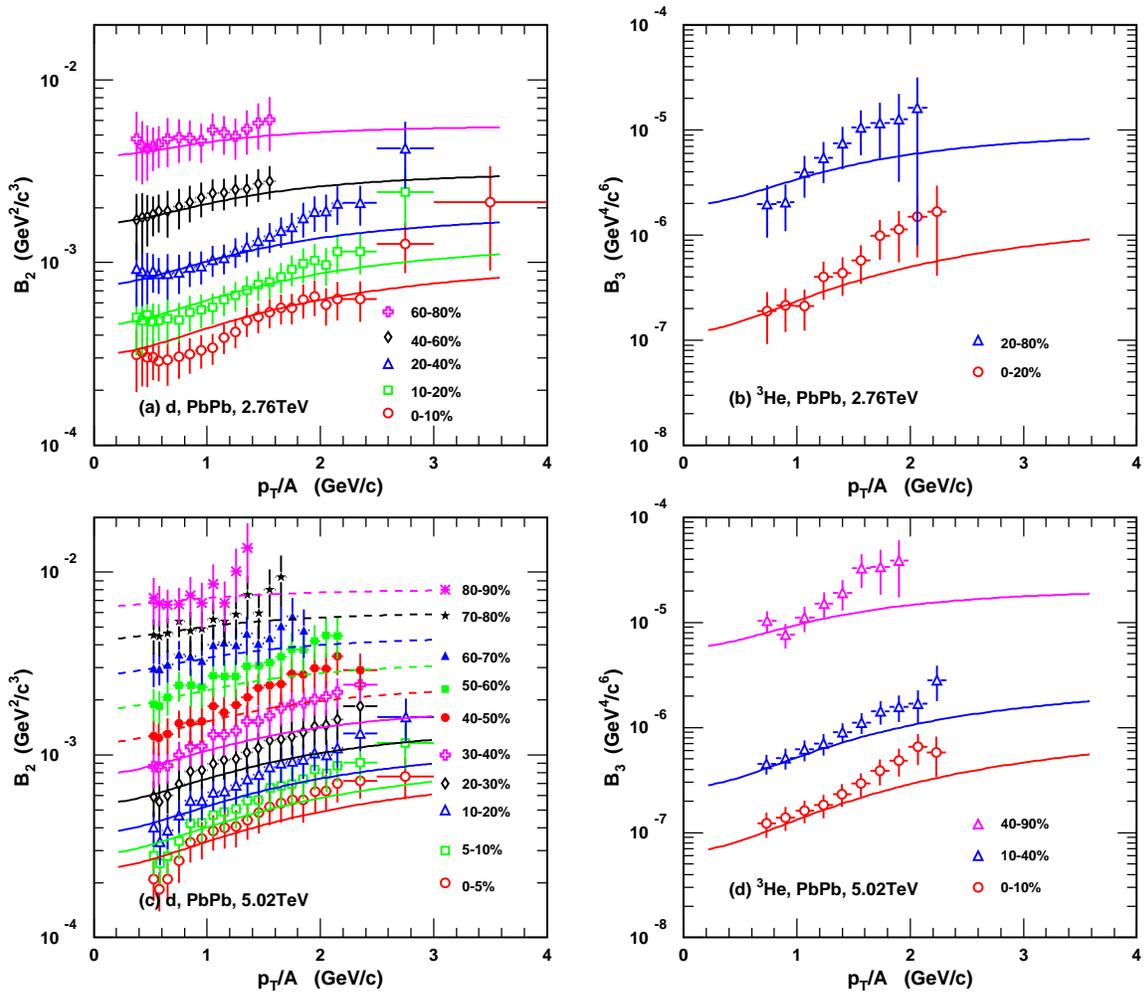}\\
 \caption{(Color online) (a) $B_2$ of deuterons and (b) $B_3$ of $^3$He as functions of $p_T/A$ in Pb-Pb collisions at $\sqrt{s_{NN}}=2.76$ TeV. 
 (c) $B_2$ of deuterons and (d) $B_3$ of $^3$He as functions of $p_T/A$ in Pb-Pb collisions at $\sqrt{s_{NN}}=5.02$ TeV.
 Symbols with error bars are the data~\cite{PbPb2016PRCALICE,PbPb2017EPJCALICE,PbPb5TeVthesis} and different lines are our results in different centralities.
 }
 \label{fig:BA-ptA276-502}
\end{figure*}

We finally calculate $B_2$ and $B_3$ as functions of $p_T/A$ in Pb-Pb collisions at $\sqrt{s_{NN}}=2.76$ TeV and 5.02 TeV.
We use the data of $dN_{ch}/d\eta$ in Refs.~\cite{dNchPbPb2013PRC,dNchPbPb2020PRC} to determine $R_f$.
$a$ is 0.43 for $^3$He, the same as those in p-p and p-Pb collisions,
while it is 0.51 for $d$.
Symbols with error bars in Fig.~\ref{fig:BA-ptA276-502} are the data~\cite{PbPb2016PRCALICE,PbPb2017EPJCALICE,PbPb5TeVthesis}, and different lines are our theoretical results.
From Fig.~\ref{fig:BA-ptA276-502}, one can see that the rising behaviors of the data can be described by our results and this rising trend becomes stronger from peripheral to central collisions.
This rising trend as the function of the transverse momentum can be naturally explained in our model by setting nucleon coalescence criteria in the rest frame of the nucleon-pair (three nucleon cluster) rather than in the laboratory frame.
Recalling that $B_A$ presents the probability of nucleons combining into light nuclei, when transformed from the laboratory frame to the nucleon-pair rest frame, the coordinate space which the nucleons in is Lorentz contracted.
So their relative distance becomes smaller and it becomes easier for them to coalescence into light nuclei.
The larger transverse momentum means larger velocity and also means stronger contraction.
This leads to larger coalescence probability.

At the end of this section, we want to point out that $R_f$ is parameterized to be $p_T$-independent to give numerical results in this paper.
From the latest experimental measurements for femtoscopic correlations of particle pairs (two pions, two kaons, and two (anti)protons), 
one sees $R_f$ exhibits a decreasing trend as a function of $p_T$~\cite{RinvPbPb2015PRC}.
Our Eqs.~(\ref{eq:B2pT}) and (\ref{eq:B3pT}) clearly show that $B_2$ and $B_3$ would increase stronger if $R_f$ decreases as a function of $p_T$, especially in central heavy ion collisions.
This would improve our numerical results in Pb-Pb collisions in Fig.~\ref{fig:BA-ptA276-502}.
But due to very limited $p_T$ range and large error bars of $R_f$ obtained from two (anti)proton correlation measurements,
it is hard to quantitively study effects on $B_A$ factors resulted from $p_T$ dependence of $R_f$ currently.

\section{summary}

Inspired by the interesting behaviors of the coalescence factors $B_A$'s of light nuclei measured experimentally,
we studied the momentum dependence of the production of deuterons and helions in high energy collisions in the framework of the nucleon coalescence/recombination.
We derived the momentum spectra for $d$ and $^3$He.
In order to get intuitionistic expressions for momentum dependence of light nuclei, in particular, those for $B_A$ factors, 
we took a few assumptions and/or approximations such as the factorization of coordinate and momentum dependencies of the kernel functions and the normalized joint nucleon distributions.
We obtained simple formulas of the momentum spectra of $d$ and $^3$He,
and in particular, we gave analytic expressions for momentum dependent $B_A$'s and discussed their properties as functions of the collision system size as well as the light nuclei size and momentum.

We applied the deduced results to the midrapidity regions of p-p, p-Pb and Pb-Pb collisions at the LHC.
We reproduced the rapidly decreasing behaviors of the data of $B_2$ and $B_3$ as the function of $dN_{ch}/d\eta$.
In central and semi-central Pb-Pb collisions at $dN_{ch}/d\eta>100$, we found that the effective radius of the system at $d$ freeze-out was about 20\% larger than that at $^3$He freeze-out.
Since the larger radius usually means the later time during system expansion evolution, our results thus indicated later freeze-out for $d$ compared to $^3$He in Pb-Pb collisions. 
Furthermore, we gave natural explanations for the obvious growth of $B_A$ against $p_T$ for all centralities in Pb-Pb collisions and relatively weak $p_T$ dependencies of $B_A$'s in p-p and p-Pb collisions at the LHC.

At last, we want to discuss that the present paper focuses on common features of the coalescence/recombination mechanism in describing the production of light nuclei in different collision systems,
and the dynamical details in the coalescence process are not included.
We try our best to intuitively present effects of kernel functions in the rest frame of the nucleon cluster and their interesting influences on $B_A$'s at different $p_T$ bins in p-p, p-Pb and Pb-Pb collisions.
We have taken the coordinate and momentum factorization assumption which means that the flow effect is ignored.
The flow in different collision systems is very different, and its effect on light nuclei production, especially in relativistic heavy ion collisions, is another interesting issue and deserves to be further considered carefully.

\section*{Acknowledgements}

We thank Xiao-Feng Luo and Lie-Wen Chen for helpful discussions.
This work was supported in part by the National Natural Science Foundation of China under Grant No. 11975011, the Natural Science Foundation of Shandong Province, China, under Grants No. ZR2020MA097, No. ZR2019YQ06 and No. ZR2019MA053, and Higher Educational Youth Innovation Science and Technology Program of Shandong Province under Grants No. 2020KJJ004 and No. 2019KJJ010.

\end{document}